# An Improved Method for the Fitting and Prediction of the Number of COVID-19 Confirmed Cases Based on LSTM

Bingjie Yan[1], Xiangyan Tang[1,*], Boyi Liu[2,*], Jun Wang[1], Yize Zhou[1], Guopeng Zheng[1], Qi Zou[1], Yao Lu[1], Wenxuan Tu[3] and Naixue Xiong[4]

**Abstract:** New coronavirus disease (COVID-19) has constituted a global pandemic and has spread to most countries and regions in the world. Through understanding the development trend of confirmed cases in a region, the government can control the pandemic by using the corresponding policies. However, the common traditional mathematical differential equations and population prediction models have limitations for time series population prediction, and even have large estimation errors. To address this issue, we propose an improved method for predicting confirmed cases based on LSTM (Long-Short Term Memory) neural network. This work compares the deviation between the experimental results of the improved LSTM prediction model and the digital prediction models (such as Logistic and Hill equations) with the real data as reference. Furthermore, this work uses the goodness of fitting to evaluate the fitting effect of the improvement. Experiments show that the proposed approach has a smaller prediction deviation and a better fitting effect. Compared with the previous forecasting methods, the contributions of our proposed improvement methods are mainly in the following aspects: 1) we have fully considered the spatiotemporal characteristics of the data, rather than single standardized data; 2) the improved parameter settings and evaluation indicators are more accurate for fitting and forecasting. 3) we consider the impact of the epidemic stage and conduct reasonable data processing for different stage.

**Keywords:** COVID-19, LSTM model, predictive analysis

# 1 Introduction

At the beginning of 2020, COVID-19 was spread worldwide [World Health Organization (2020)]. The analysis and prediction of COVID-19 is a vital prerequisite for formulating infectious disease prevention and control strategies. The prediction of infectious diseases is an important task in health work, which will detect the development trend of the disease early and increase the predictability of the epidemic prevention work [Wang Xinzhi; Liu Yi; Zhang Hui; Ma Qiuju; Cao Zhidong. (2018)]. It plays an essential rule in disease prevention, treatment and health decision-making.

---

[1] School of Computer Science and Cyberspace Security, Hainan University, Haikou, 570228, China.

[2] University of Chinese Academy of Science, Beijing, China.

[3] National University of Defense Technology, Changsha, China.

[4] Department of Mathematics and Computer Science, Northeastern State University, Tahlequah, OK, USA

*Corresponding Author: Xiangyan Tang. Email: txy36@163.com. Boyi Liu. Email: by.liu@ieee.org.

The confirmed cases prediction model for COVID-19 is still in the research and exploration stage. Among them, how to use machine learning to predict the effect still needs more experiments to verify.

The number of infections and deaths of infectious diseases is a set of time series. There are many methods for time series prediction [Cheng, J.; Xu, R. M.; Tang, X. Y.; Sheng, V. S.; Cai, C. T. (2018)], [Oh Byoung Doo; Song Hye Jeong; Kim Jong Dae; Park Chan Young; Kim Yu Seop. (2019)]. As a special RNN model, LSTM has always been effective in time-series data prediction.

RNN can capture the dynamics of the sequence through loops in the network of nodes. The recursive network retains a state that can represent information from any long context window. In the past, it is challenging to train RNN because the training of RNN model usually contains millions of parameters. However, in recent years, with the development of network architecture, optimization technology and parallel computing, it is possible to train RNN models on a large scale. Among them, the models based on the long-term short-term memory system (LSTM) and Bidirectional RNN (BRNN) have a great breakthrough in performance and can complete a variety of tasks.

The feedback connection of LSTM makes the neural network have internal or short-term memory. Such features are suitable for processing sequence-related problems: such as speech classification, prediction and generation, image subtitles, language translation, and handwriting recognition.

In view of the current COVID-19 infection status, this paper proposes a LSTM-based COVID-19 confirmed cases prediction model. The core model of this paper is based on the LSTM model. The outstanding advantage of the LSTM model is that it can effectively capture long-term time dependencies. The input of the model is mainly the number of diagnoses, the number of deaths, whether the city is closed, and the features included are the number of newly diagnosed, the number of deaths, the growth rate of the number of diagnoses, the growth rate of deaths, etc. The final output is the predicted number of people who get infected.

In this paper, we propose an improved method. This method has a better fitting effect for regions with a large population base, and the prediction effect is more accurate than the traditional Logistic and Hill Equation population prediction algorithms.

**2 Related Work**

This paper is based on the background data of the number of confirmed diagnoses and deaths of new coronavirus (COVID-19) outbreaks in Wuhan, China, in December 2019. We reviewed the traditional population growth prediction model, infectious disease model, the number prediction of machine learning applied in practice, and the recent research and development prediction model of COVID-19. Then, an improved prediction approach based on LSTM is proposed in the work.

## 2.1 Population growth prediction model

Since the Malthus population model was proposed, it has been widely used in various fields. Later, related scholars added the population coefficient to the Malthus model to realize the improvement of the model, and proposed the Logistic model. This model makes the related population prediction more accurate and effective. For the prediction of regional population size and age structure, Shorokhov, S. I. used the matrix form to realize population prediction through the comparison of immigration inflow [Shorokhov, S. I. (2014)]. Similarly, the use of different mathematical models can also predict the number of world population. Yuri S. Popkov and others used the entropy estimation algorithm to construct a random prediction model [Popkov Y S, Dubnov Y A, and Popkov A Y. (2016)]. The problem with the traditional Logistic regression model is that when there are many feature quantities, the accuracy of its prediction will decrease. Therefore, there are not suitable for scenarios containing a large number of multi-type features or variables. However, we propose a prediction method using the LSTM network model in machine learning which has a good advantage in this filed.

## 2.2 Infectious Disease Model

According to the population growth prediction model, Kermack proposed the SIR epidemic model. After that, the SIR epidemic model has been well applied. Ning, Z et al. predicted the case of infectious diseases through grey and differential equation models [Ning, Z. and Lin, L. (2014)]. This model is mainly used in epidemiology. A common situation is to explore the risk factors of a disease or predict the probability of a disease based on risk factors and so on. Paul D. Haemig used decades of database to build mathematical models and realized the prediction of the number of patients with encephalitis next year [Haemig P D, Sjöstedt de Luna S, Grafström A et al. (2011)]. In terms of infectious diseases related to this article, there are also many mathematical models used to predict. A. Gray used stochastic differential equations to improve the SIS epidemic model [Gray A, Greenhalgh D, Hu L et al. (2011)] and established a new SIR model to predict the development trend of avian influenza and human influenza [Iwami S, Takeuchi Y, and Liu X. (2007)]. In recent years, infectious disease models have been continuously improved and developed. Infectious disease models are applied to the prediction and prevention of medical places and have played a huge role in promoting it. There is a combination of approximate Bayesian algorithm and infectious disease model to calculate risk probability [Minter A and Retkute R. (2019)]. Li Q combined with evolutionary game theory to realize the infectious disease prediction model with vaccination strategy, and achieved good results in practical application [Li Q, Li M C, Lv L et al. (2017)].In response to the development of national conditions, relevant scholars have also proposed the practical comparison of real-time prediction of endemic infectious diseases based on LASSO models in different countries [Chen Y, Chu C W, Chen M I C et al. (2018)]; In addition, there are many practical models. On the one hand, the comparison of basic models enables the selection, trade-off and comparison of infectious disease models [Funk S and King A A. (2020)]. On the other hand, the researchers combined the local ILI incidence and used a dynamically calibrated compartment model for real-time analysis and real-time prediction of influenza outbreaks in Belgium [Miranda G H B, Baetens J M, Bossuyt N et al. (2019)]. Although the SIR epidemic model predicts good results, there has a big problem. This model is based on differential equations, so the process of solving equations is very complicated. Another drawback is the prediction results are very sensitive to the initial value conditions, leading that the robustness of the model is week. In this respect, we use the LSTM model's own reverse error propagation characteristics to train the data.

### *2.3 Quantity prediction of machine learning applied to practice*

The traditional prediction methods are based on the mathematical model. Nowadays, Machine Learning prediction has more and more widely used. The combination of Machine Learning and traditional forecasting models for quantitative forecasting has mushroomed. At present, there are data processing methods using hierarchical learning methods to predict the citation times of future papers [Chakraborty T, Kumar S, Goyal P et al. (2014)]. Some people have also used machine learning to realize the measurement of content and literature, and predict the citation time of biomedical literature [Fu L and Aliferis C. (2010)]. In addition, artificial neural networks have also been used to predict populations [Folorunso O, Akinwale A T, Asiribo O E et al. (2010)]. At the same time, because of the powerful generalization ability of ML, the traffic flow prediction model based on Machine Learning has been widely used recently. The existing method introduces the maximum entropy Kalman filter to achieve traffic flow prediction [Cai L, Zhang Z and Yang J et al. (2019)]. These studies use deep learning neural networks to achieve traffic flow prediction. For example, they are the methods to realize multi-intersection traffic flow prediction learning [Shen Z, Wang W, Shen Q et al. (2019)] and hybrid traffic flow prediction method based on multi-mode deep learning [Du S, Li T, Gong X et al. (2018)]. In addition, Machine Learning has also been commonly used in infectious disease model prediction. Hani M. Aburas used a neural network model to predict the proportion of patients diagnosed with dengue fever [Aburas H M, Cetiner B G and Sari M. (2010)]. At present, machine learning prediction models are often used to predict traffic flow and other situations. However, there is relatively little research on the prediction number of people. Our research applies the LSTM model to the number prediction and finds that the results are relatively good.

### *2.4 Prediction models for recent studies on COVID-19*

For the recent research and development of COVID-19 and its impact, many researchers have made relevant prediction models based on the data. Zifeng Yang et al. analyzed the epidemic prevention measures of the Chinese government, and predicted the epidemic situation of COVID-19 [Yang Z, Zeng Z, Wang K et al. (2020)]. They used the improved SEIS model and the AI network trained by the SARS data in 2003, and predicted the next epidemic situation in China. The facts proved that their prediction results were relatively good. By using the real-time data on the website, Dong E et al. built a web-based interactive dashboard that can track COVID-19 in real time using the characteristics of Internet data [Dong E, Du H and Gardner L. (2020)]. In addition, there is a dynamic stochastic general equilibrium (DSGE) model that uses data randomization to assess the impact of coronavirus on the tourism industry and achieves trend prediction [Yang Y, Zhang H and Chen X. (2020)]. Although there have been many studies on the prediction model of COVID-19 recently, we first used the LSTM model to consider the end of the epidemic. The LSTM model performs forward time memory processing on the collected large amount of data. And it considers actual factors and characteristic conditions, and outputs data associated with the factors.

## 3 Methodology

In this session, we will specifically introduce our data processing method, the construction of the LSTM model, and the improved method.

### *3.1 Data selection and processing*

Novel coronavirus pneumonia COVID-19 is a highly contagious virus. We hope to use the data on the number of confirmed cases of the previous days to predict the growth trend of the number of confirmed cases in the following days. At the same time, whether the national government takes favorable measures to cut off the spread of the virus is also an important factor, which will affect the time when the inflection point appears and the time when the virus continues to spread. Therefore, we use the data onto the number of daily diagnosing as different countries and the time when the country takes similar "the curfew" measures as the input of the model data.

To increase the learning efficiency and the features extracting efficiency, it is necessary to preprocess the data and display more features to facilitate model learning. We take the cumulative number of diagnosed, the number of newly diagnosed, the cumulative growth rate of daily diagnoses, and whether the government has closed the city as the input of model learning. Simultaneously, the number of newly confirmed cases of the next day is used as an output rather than the cumulative diagnosis of the next day. Because the numerical range of the number of newly confirmed cases will be relatively small, which is convenient for the evaluation and comparison of the loss function. For the first two items, you should not simply use MinMaxScaler for standardization, because the value of Max will be broken. So we need to manually set the Max value so that the input data is between $[0,1]$, which can predict the next few days of data perform better, you can directly use 0 and 1 to express the two options of whether to close the city.

### *3.2 Long Short Term Memory networks (LSTM)*

Long Short Term Memory networks are a special kind of Recurrent Neural Networks. It has a better effect on time series prediction. There is an incubation period for the new coronavirus. Using LSTM for time series prediction may find the influencing factors of potential cases.

Entering LSTM will first pass a forgetting gate, and pass a sigmoid layer which is also called forgetting gate layer. Take the current state $h_{t-1}$ and the new input $x_t$ as the input of this layer, and the output is a value of $[0,1]$. Indicating whether a kind of forget to store in the previous cell The decision on $C_{t-1}$, 0 means completely forgotten, 1 means completely reserved, which is

$$f_t = \sigma(w_f \cdot [h_{t-1}, x_t] + b_f) \tag{1}$$

Next, you will enter the LSTM input gate. In this step, we determine the vector to be updated and create a vector update candidate value of a tanh activation function. It can be expressed by the following formula:

$$i_t = \sigma(w_i \cdot [h_{t-1}, x_t] + b_i) \tag{2}$$

$$\tilde{C}_t = \tanh(w_C \cdot [h_{t-1}, x_t] + b_C) \tag{3}$$

Then you can get the status in the storage cell as

$$C_t = f_t \cdot C_{t-1} + i_t \cdot \tilde{C}_t \tag{4}$$

Finally, the last decision output gate will be used to decide what to output and update the status in a similar way as the previous two gates.

$$o_t = \sigma(w_o \cdot [h_{t-1}, x_t] + b_o) \tag{5}$$

$$h_t = o_t \cdot \tanh C_t \tag{6}$$

In this way, we get a prediction of future output.

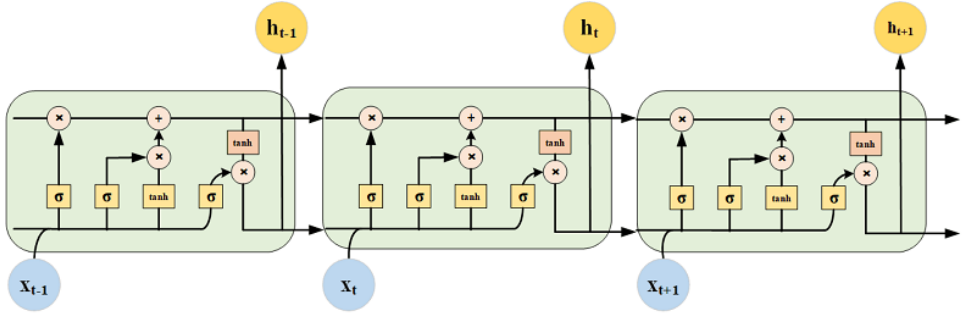

**Figure 1:** Four interacting layers in LSTM

*3.3 Proposed Improved Method*

We found that the results of the data fitting phase are ideal for regions with few cases, but for regions with a large number of cases, such as Hubei, China, the results of the fitting experiment are biased. In this respect, we propose the following methods to improve the results of the fitting phase which can also make the prediction results better.

This type of problem mainly occurs in regions where the epidemic is serious. Because the base of the number of people infected with the virus is large, a relatively large number of new people will be obtained when there should be a small number of new people in the later stage of the epidemic. In response to this problem, we propose the following methods to improve the prediction results of the ordinary LSTM model.

First of all, in the later stage of the epidemic, the prevention and control of the epidemic is relatively strong, and the data fluctuation is small. Simultaneously, the number of infected people and the number of diagnosing can be considered to be roughly equal, because the diagnosed patients will be isolated and can be considered no longer contagious. Therefore, this stage should be carried out under the conditions of a small population base.

So we use the standard deviation of last *n* days as the judgment condition to adjust the parameters for the number of confirmed cases. First, the data to be input is normalized by MinMaxScaler based on itself, to unify the standard deviation judgment standard, and then calculate the variance $\delta^2$. If the variance $\delta^2$ is greater than the critical value $\alpha$, that is $\delta^2 > \alpha$, then the data is reduced based on the population base method, divided by the logarithm based on $\beta$, and $\beta$ can be calculated by simple adjustment. The specific improved algorithm is as follows:

**Algorithm 1**

**Input:** Original data *x*, *n* is the length of the data used for conditional judgment, $\alpha$ is conditional judgment boundary, $\beta$ is the base of the calculated zoom logarithm.
**Output:** Data after condition judgment and processing
1: $x' = x[-n:]$
2: $x'$=MinMaxScaler(x')
3: $\delta^2 = \frac{1}{n} \sum_{i=1}^{n}(x'_i - \bar{x'})^2$
4: **if** ($\delta^2 > \alpha$) **then**
5:     $x = \log_\beta x$
6: **end if**

### *3.4 Model*

Because of the existence of an incubation period and asymptomatic carriers, the Novel coronavirus pneumonia has many potential factors. However, discovering these potential effects requires many characteristics. The results obtained only by fitting the LSTM to the trend may not be ideal. So we go through a fully connected neural network to further extract the features of the data as the input of the LSTM layer at first. The LSTM layers output predictions for different conditions of various characteristics according to various characteristics, and then integrates predictions of various characteristics through the fully connected layer to obtain the desired result.

At the same time, the incubation period of the Novel coronavirus pneumonia can be as long as 14 days, so the time series sequences we entered should theoretically be greater than 14 days.

## 4 Experiment

In this section, we will introduce our experiments using model fitting prediction, and explain our evaluation indicators and parameter settings.

### *4.1 Dataset*

*Novel Coronavirus (COVID-19) Cases.* It is a dataset for the number of COVID-19 confirmed cases, deaths and recovered operated by the Johns Hopkins Center for Systems Science and Engineering (JHU CSSE) with the support of the ESRI Living Atlas team and the Johns Hopkins University Applied Physics Laboratory (JHU APL). It's available online at https://github.com/CSSEGISandData/COVID-19

*COVID-19 Lockdown dates by country.* It is a table of the lockdown dates of countries or provinces collected by jcyzag from the internet and published on the kaggle dataset. It's available online at https://www.kaggle.com/jcyzag/covid19-lockdown-dates-by-country.

### *4.2 Evaluation Metric*

#### *4.2.1 Goodness of fit*

The goodness of fit of a statistical model describes how well it fits a set of observations. The scale of goodness of fit usually summarizes the difference between the observed values and the expected values under the model.

Calculation method. Note the mean value of the values to be fitted is $\bar{y}$, the expected value is $\hat{y}_i$, and the fitting result is $y_i$.

Total sum of squares SST is:

$$SST = \sum_{i=1}^{n}(y_i - \bar{y})^2 \tag{7}$$

Regression sum of squares SSR is:

$$SSR = \sum_{i=1}^{n}(\hat{y}_i - \bar{y})^2 \tag{8}$$

From this calculation, the goodness of fit can be obtained.

$$R^2 = \frac{SSR}{SST} = \frac{\sum_{i=1}^{n}(y_i - \bar{y})^2}{\sum_{i=1}^{n}(\hat{y}_i - \bar{y})^2} \tag{9}$$

#### *4.2.2 Deviation Rate*

Note that the expected value is $\hat{y}_i$, and the fitted value is $y_i$, then

$$\text{Deviation Rate} = \frac{1}{n}\sum_{i=1}^{n}\left|\frac{\hat{y}_i - y_i}{y_i}\right| \times 100\% \tag{10}$$

### *4.3 Parameters*

#### *4.3.1 Time series length*

We set a training window of the training data and get a time series of data. Our model set a time-series sequence of up to 21 days for training, with the aim of discovering more carriers' influence on virus transmission during the incubation period.

*4.3.2 Learning Rate*

After normalization, the absolute values of all input and output data values are small, so a small learning rate is required for the learning progress. As the number of training epochs increases, the learning rate should also decrease, so that the data can be better fitted.

*4.3.3 Forecast days*

In order to show the results of data fitting and prediction, we use the trained model to predict the number of confirmed cases in the next 7 days.

*4.3.4 Model*

Our model is built according to the following steps. First, the input data extraction features were put into the LSTM to obtain several time series outputs. After getting the output of the LSTM network, integrate and merge features in the last part. The model structure is shown in Fig 2.

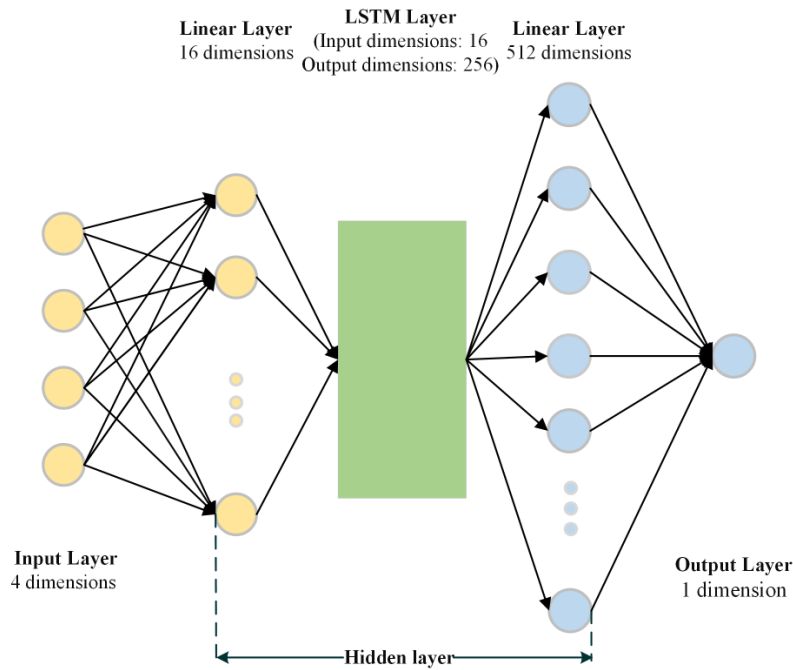

**Figure 2:** The model used in the experiment

*4.4 Experiment Result*

This part shows the results of our experiments. We will compare our improved LSTM prediction results with the results of the number prediction model such as Logistic and Hill Equations. At the same time, the fitting results of our improved fitting algorithm and the unimproved LSTM algorithm will also be compared.

During the training process, learning rate decreases with the increase of epoch, and the loss convergence during the training process is shown in Fig 3.

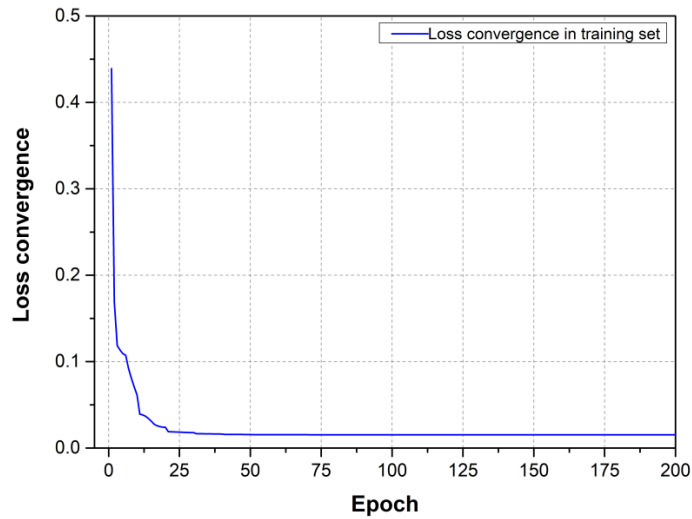

**Figure 3:** Loss convergence in the training set

We selected the case data of Tianjin, China, Hong Kong, South Korea and Italy for fitting and prediction after the process of model training. The experimental results are shown in the Fig 4.

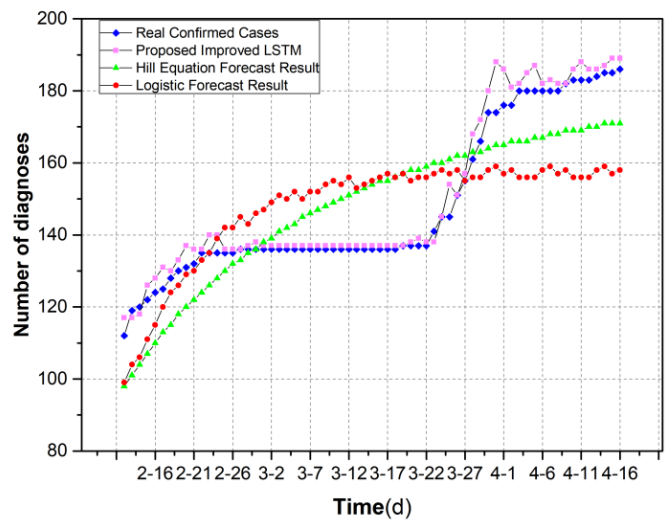

**Figure 4(a):** Fit and forecast of the number of confirmed cases in Tianjin, China.

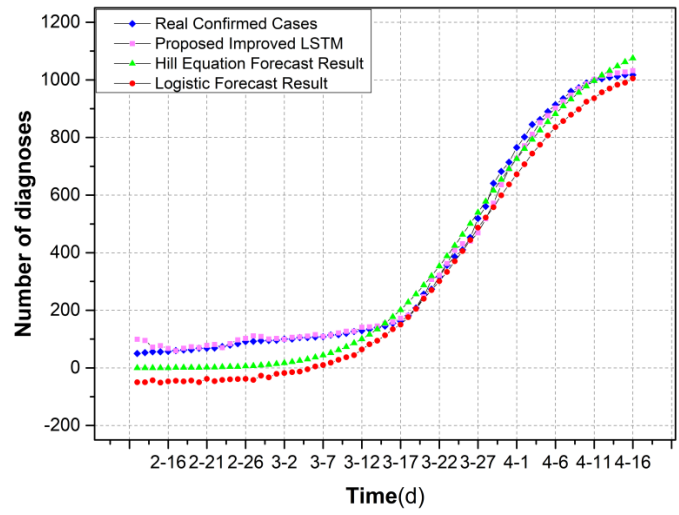

**Figure 4(b):** Fit and forecast of the number of confirmed cases in Hongkong, China.

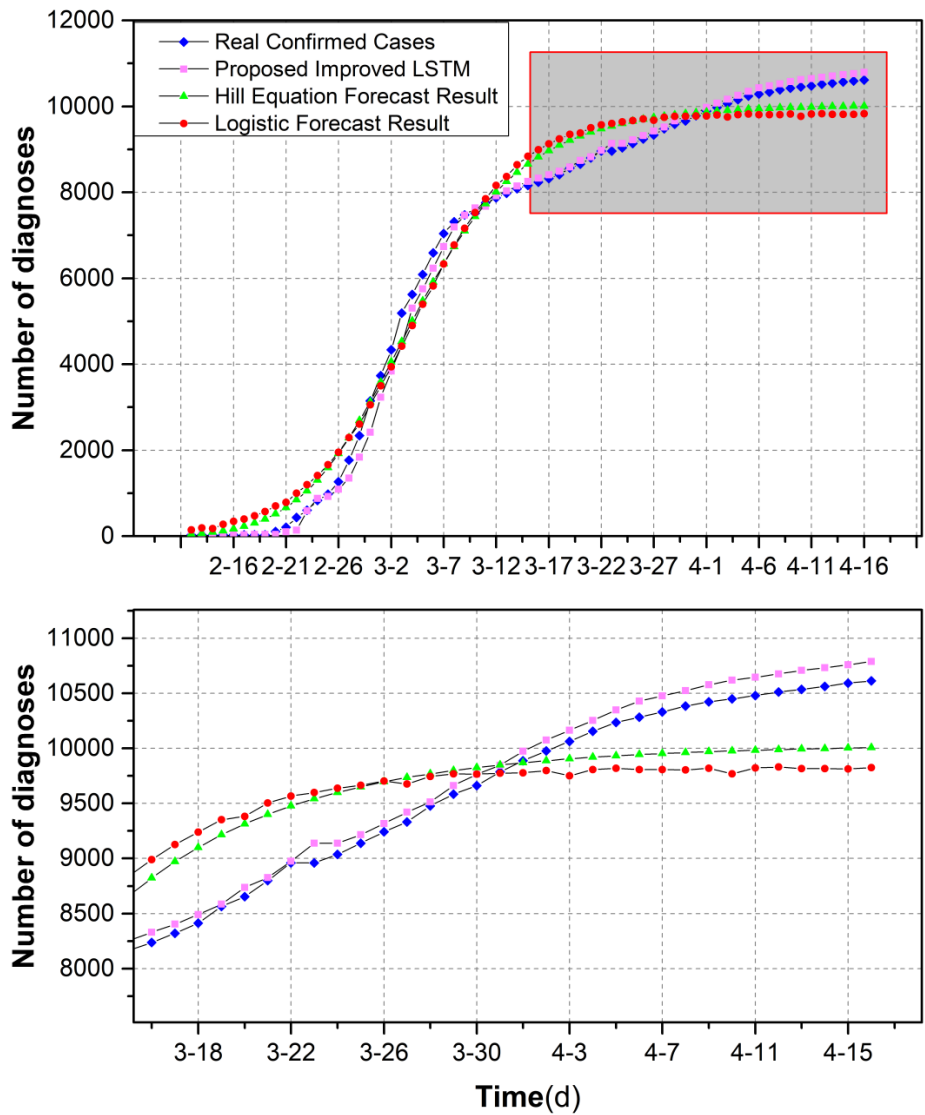

**Figure 4(c):** Fit and forecast of the number of confirmed cases in South Korea.

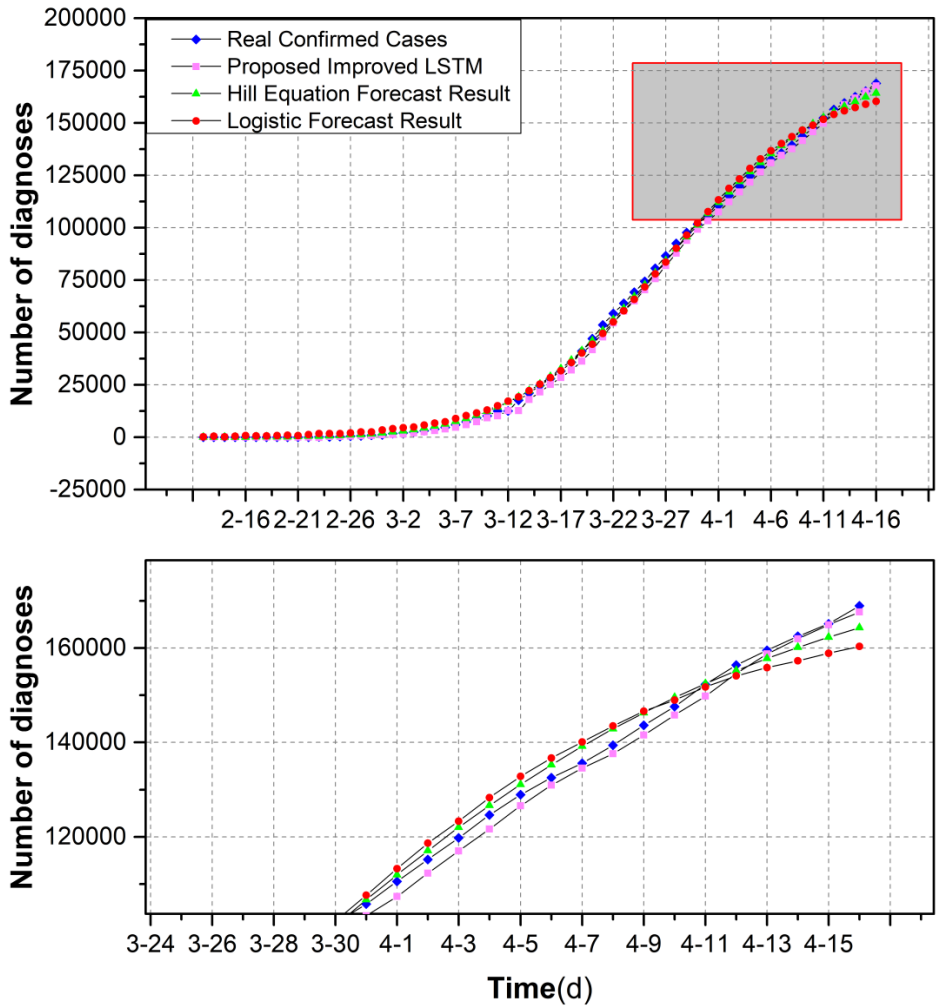

**Figure 4(d):** Fit and forecast of the number of confirmed cases in Italy.

We rounded up the specific prediction results obtained by the proposed improved LSTM algorithm, Logistic algorithm, and Hill Equation algorithm for each city shown in the figure to take integers and real case data, and record them in the Tab 1.

**Table 1(a):** Forecast of the number of confirmed cases in Tianjin,China.

| Date | Real | **LSTM** | Logistic | Hill Equation |
|---|---|---|---|---|
| 2020/4/10 | 183 | **186** | 156 | 169 |
| 2020/4/11 | 183 | **188** | 156 | 169 |
| 2020/4/12 | 183 | **186** | 156 | 170 |
| 2020/4/13 | 184 | **186** | 158 | 170 |
| 2020/4/14 | 185 | **187** | 159 | 171 |
| 2020/4/15 | 185 | **189** | 157 | 171 |
| 2020/4/16 | 186 | **189** | 158 | 171 |

**Table 1(b):** Forecast of the number of confirmed cases in Hongkong,China.

| Date | Real | **LSTM** | Logistic | Hill Equation |
|---|---|---|---|---|
| 2020/4/10 | 989 | **1014** | 924 | 978 |
| 2020/4/11 | 1000 | **1030** | 936 | 997 |
| 2020/4/12 | 1004 | **1040** | 957 | 1016 |
| 2020/4/13 | 1009 | **1042** | 970 | 1032 |
| 2020/4/14 | 1012 | **1046** | 983 | 1048 |
| 2020/4/15 | 1017 | **1049** | 990 | 1062 |
| 2020/4/16 | 1017 | **1054** | 1005 | 1075 |

**Table 1(c):** Forecast of the number of confirmed cases in South Korea.

| Date | Real | **LSTM** | Logistic | Hill Equation |
|---|---|---|---|---|
| 2020/4/10 | 10450 | **10618** | 9770 | 9977 |
| 2020/4/11 | 10480 | **10645** | 9823 | 9983 |
| 2020/4/12 | 10512 | **10675** | 9832 | 9989 |
| 2020/4/13 | 10537 | **10708** | 9817 | 9994 |
| 2020/4/14 | 10564 | **10733** | 9817 | 9999 |
| 2020/4/15 | 10591 | **10760** | 9812 | 10003 |
| 2020/4/16 | 10613 | **10788** | 9826 | 10007 |

**Table 1(d):** Forecast of the number of confirmed cases in Italy.

| Date | Real | **LSTM** | Logistic | Hill Equation |
|---|---|---|---|---|
| 2020/4/10 | 147577 | **145820** | 148959 | 149502 |
| 2020/4/11 | 152271 | **149824** | 151754 | 152473 |
| 2020/4/12 | 156363 | **154606** | 154093 | 155225 |
| 2020/4/13 | 159516 | **158744** | 155852 | 157770 |
| 2020/4/14 | 162488 | **161923** | 157324 | 160118 |
| 2020/4/15 | 165155 | **164938** | 158871 | 162282 |
| 2020/4/16 | 168941 | **167638** | 160330 | 164274 |

We perform statistical analysis on the prediction results and calculate the deviation value. It can be seen that the improved LSTM model prediction results are better than the traditional Logistic and Hill Equation prediction models in most cases. The average error can be controlled within 2%, which is better than the Logistic and Hill Equation performance.

**Table 2:** Deviation rate of forecast results in the above regions

|  | **Improved LSTM** | Logistic | Hill Equation |
|---|---|---|---|
| Tianjin, China | **1.70772%** | 14.66224% | 7.60182% |
| Hongkong, China | **0.86212%** | 4.03131% | 2.65315% |
| South Korea | **1.59999%** | 5.14392% | 6.84572% |
| Italy | **0.80794%** | 2.44353% | 1.31715% |

Since most of the improved algorithms are suitable for data fitting at the end of epidemic prevention and control, and now foreign epidemic situations other than China are still in the outbreak period, they are not applicable. Therefore, we selected three provinces in China, Hubei Province, Henan Province, and Zhejiang Province, where the epidemic situations are more serious, and the base of infection cases is larger. The figures show the experimental results we obtained. It can be seen that the adjusted fitting results are closer to the real data. Although there may be a large error during the outbreak of the epidemic, the prediction result has a small error at the end of the outbreak.

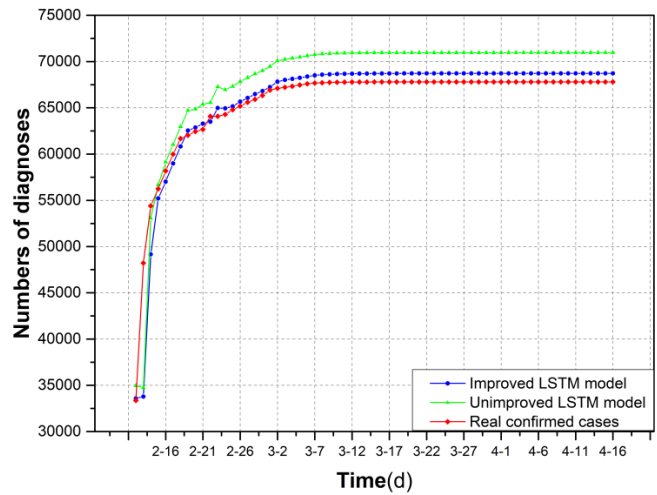

**Figure 5(a):** The fitting result of the number of people diagnosed in Hubei, China

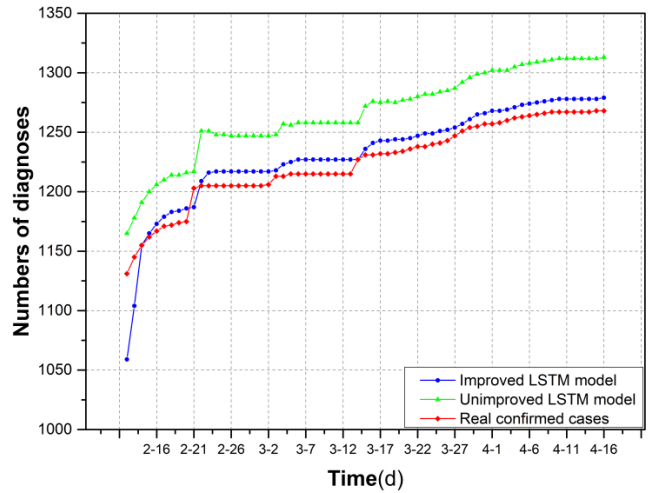

**Figure 5(b):** The fitting result of the number of people diagnosed in Zhejiang, China.

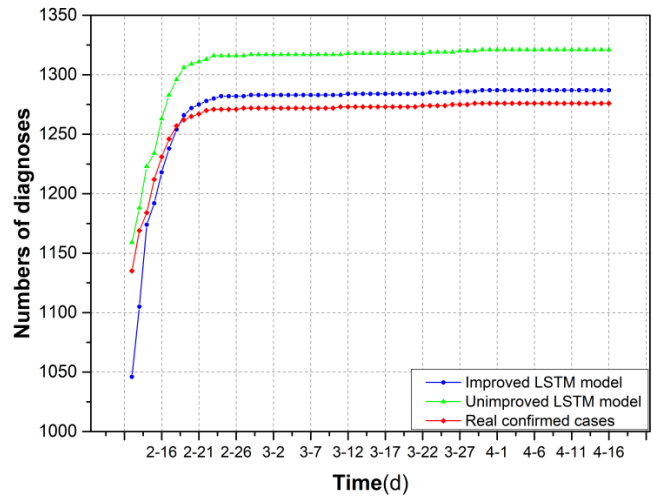

**Figure 5(c):** The fitting result of the number of people diagnosed in Henan, China

We calculated the goodness of fit for the fitting results of the two models before and after improvement, as shown in Tab 3. The prediction result of the improved model will be closer to the true value, effectively reducing the late prediction error due to the large case base.

**Table 3:** Comparison of goodness of fit of LSTM model before and after improvement

|                  | **Improved LSTM** | Unimproved LSTM |
|------------------|-------------------|-----------------|
| Hubei, China     | **0.6332**        | 0.5382          |
| Zhejiang, China  | **0.6528**        | 0.3776          |
| Henan, China     | **0.3747**        | 0.2307          |

The relative error of the number of daily cases after fitting by two methods is shown in the Fig 6.

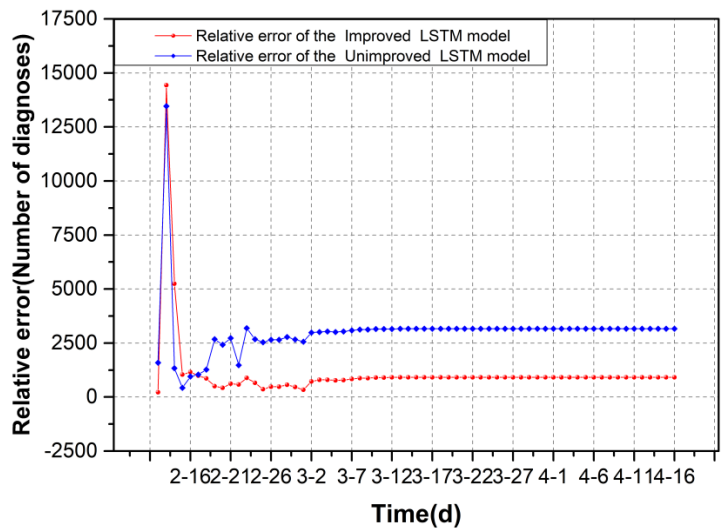

**Figure 6(a):** The relative error of fitting results of case data in Hubei, China.

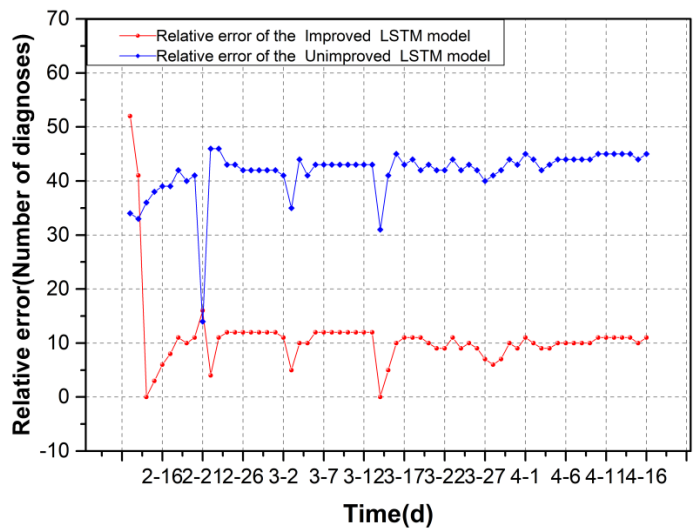

**Figure 6(b):** The relative error of fitting results of case data in Zhejiang, China.

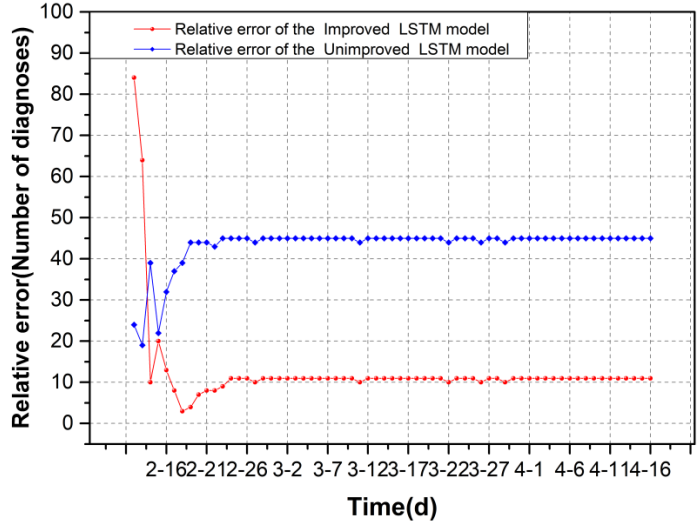

**Figure 6(c):** The relative error of fitting results of case data in Henan, China.

## 5 Conclusion

To address the problem of deviation and accuracy in predicting the number of confirmed cases in traditional methods, we propose an improved method based on LSTM neural network. The prediction of the diagnosis number of new coronavirus can be regarded as time series prediction, and the LSTM model has a good effect on time series prediction. Because of the incubation period of new coronaviruses, time series prediction using LSTM may also find the influencing factors of potential cases.

First of all, we use the 21-day case data of various countries and regions provided on the website, and use the cumulative number of diagnoses, the number of new diagnoses, the cumulative growth rate of daily diagnoses, and whether the government closes the city as the model input and set parameters MinMaxScaler standardized processing. Through a fully connected neural network, the features of the data are further extracted as input to the LSTM layer. Build an LSTM neural network, and set the length of the time series, the learning rate, and the number of days to predict. In view of the huge data in the training set, to reduce the fitting deviation, we improved the LSTM model, used the unified standard deviation as the judgment standard, reduced the data using the method based on the overall cardinality, and adjusted the data. This method creates a tanh activation function update vector, obtains the status in the storage unit, and determines the output and update status content through the decision output gate, which can output the predicted number of newly diagnosed cases in the next day. Besides, according to the actual situation of the actual epidemic situation, we selected data from several countries and regions, and compared the deviation between the experimental results of the improved LSTM prediction model and the digital prediction model (such as Logistic and Hill equation) and the actual results. And introduce the fitting result of the LSTM algorithm before and after the improvement. Experiments show that the improved LSTM model has a smaller prediction deviation and better fitting effect.

**Funding Statement:** This work was supported by the Hainan Provincial Natural Science Foundation of China [2018CXTD333, 617048]; National Natural Science Foundation of China [61762033, 61702539]; Hainan University Doctor Start Fund Project [kyqd1328]; Hainan University Youth Fund Project [qnjj1444]; Ministry of Education Humanities and Social Sciences Research Program Fund Project [19YJA710010]; the Opening Project of Shanghai Trusted Industrial Control Platform.

**Conflicts of Interest:** The authors declare that they have no conflicts of interest to report regarding the present study.